# Case Tool: Fast Interconnections with New 3-Disjoint Paths MIN Simulation Module

Ravi Rastogi, Amit Singh and Nikhil Singhal
Department of CSE and IT, Jaypee University of Information Technology, Waknaghat, Solan-173234, Himachal Pradesh, India

Nitin
Department of Computer Science, The Peter Kiewit Institute, College of Information Science and Technology, University of Nebraska at Omaha, Omaha-68182-0116, Nebraska, United States of America

Durg Singh Chauhan
Uttarakhand Technical University, Post Office Chandanwadi, Prem Nagar, Sudohwala, Dehradun-248007, Uttarakhand, India

## ABSTRACT
Multi–stage interconnection networks (MIN) can be designed to achieve fault tolerance and collision solving by providing a set of disjoint paths. In this paper, we are discussing the new simulator added to the tool designed for developing fault-tolerant MINs. The designed tool is one of its own kind and will help the user in developing 2 and 3-disjoint path networks. The java technology has been used to design the tool and have been tested on different software platform.

## 1. INTRODUCTION AND MOTIVATION
In a multiprocessor system, many processors and memory modules are tightly coupled together with an interconnection network. A properly designed interconnection network certainly improves the performance of such multiprocessor system. Multistage Interconnection Networks (MINs) [1-10] are highly suitable for communication among tightly coupled nodes. For ensuring high reliability in complex systems, fault tolerance is an important issue. The basic idea for fault tolerance is to provide multiple paths for a source–destination pair, so that alternate paths can be used in case of a fault in a path [1-23]. However, to guarantee 1–fault tolerance, a network should have a pair of alternate paths for every source destination pair which are disjoint in nature [1–8, 24-31].

Now-a-days applications of MINs are widely used for on-Chip communication. In past number of techniques has been used to increase the reliability and fault-tolerance of MINs, a survey of the fault-tolerance attributes of these networks is found in [1-6]. The modest cost of unique paths MINs makes them attractive for large multiprocessors systems, but their lack of fault-tolerance, is a major drawback. To mitigate this problem, three hardware options are available [1-5, 20-23]:

1. Replicate the entire network,
2. Add extra stages,
3. And /or Add chaining links.
4. Rearranging of the connection patterns with the addition or deletion of hardware links.

In addition to this, MINs can be designed to achieve fault tolerance and collision solving by providing a set of disjoint paths. Many researchers have done sufficient work on providing 1-fault tolerance to the MINs however; little attention has been paid to design the 3-Disjoint Paths Fault-tolerant MINs. We have been inspired by the work presented by the authors in [24-31].

A Multi–stage interconnection network is fully able to meet the reliability demands if it is at least one fault tolerant that is there is at least one alternative path to deal with faults or collisions. This alternative path should be disjoint in nature with the existing routing path followed so that there is no such implication that if a switch or a link fails in the existing routing path then the alternative path will also fail. Most design of Multi–stage interconnection networks do not generate at least two disjoint paths and hence are not always fault tolerant resulting in packet losses and eventual performance degradation. Hence, this approach of two disjoint paths will always guarantee a way out of the problem of faults or collisions in a network [32-34].

Whenever we want to design a interconnection network, we used to design them manually using the windows word and then hardwired them through the programming. At present, we do not have any tool through which we can develop the interconnection networks tool or this remains out of limelight therefore in this paper; we have discussed a tool designed for developing fault-tolerant multi-stage interconnection networks. The designed tool is one of its own kind and will help the user in developing 2 and 3-disjoint path networks.

The rest of the paper is as follows: Section 2 discusses the testbed and experimental setup, new modules added to the existing case tool [32-34] and algorithm supported by the screen shots and the pseudocode followed by the conclusion and references.

## 2. CASE TOOL: FAST INTERCONNECTIONS
### 2.1 Testbed and Experimental Setup
CASE stands for "Computer Assisted Software Engineering. A CASE tool is a software tool that helps software designers and developers specify, generate and maintain some or all of the software components of an application. Many popular CASE tools provide functions to allow developers to draw database schemas and to generate the corresponding code in a data description language (DDL). Other CASE tools support the analysis and design phases of software development, for example by allowing the software developer to draw different types of UML diagrams [35].

We have designed both the networks using the Fast Interconnections tool and the architectural design of the software is already published in [33-34]. We have used Eclipse, is a multi-language software development environment comprising an integrated development environment (IDE) and an extensible plug-in system. It is written mostly in Java and can be used to develop applications in Java and, by means of various plug-ins, other programming languages. The IDE is often called Eclipse JDT for Java (i.e. JDK 1.6) and IDE is running on top of the IBM System x, running with Novell's SUSE Linux Enterprise Server 11. We have used advanced java features to build our system. The most important part of the tool is designing of the components, which are used to design disjoint paths MINs. We have design them in paint and stored them in component library. We have provided the access of this component within the tool using ComponentChooser class.

## 2.2 New Module added to the Case Tool
1. Added a new 3-Disjoint Paths Multi-stage Interconnection Network Simulator,
2. Design a circuit, enter a custom path, faulty component numbers, and click on simulate button,
3. Simulation will start and the path will turn green/red, depending on the packet drop with the faulty component marked by red cross every time a packet is dropped.

## 2.3 Algorithm
Algorithmic Step 1: Get path in A.

Algorithmic Step 2: Extract individual wire numbers in path1[].

Algorithmic Step 3: Get faulty_components in B.

Algorithmic Step 4: Extract individual comp. numbers in cmp1[].

Algorithmic Step 5: Verify the correctness of the path1[] and cmp1[].

Algorithmic Step 6: For 15 seconds repeats steps 7 and 8.

Algorithmic Step 7: For each wire in path1[] display it as green color for packet transfer.

Algorithmic Step 8: For every alternate packet drop a packet at components in cmp1[], display a red cross over components in cmp1[] and wires in path1[] as red color.

Algorithmic Step 9: End Simulation.

## 2.4 Case Tool: Screen Shots

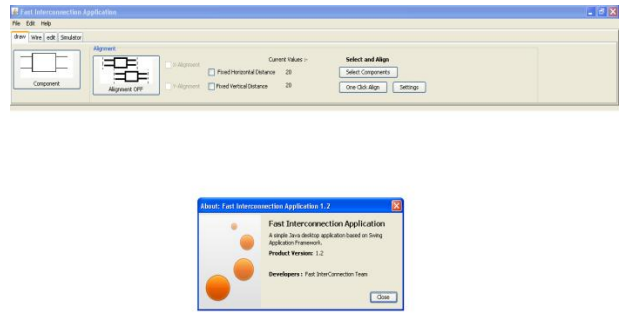

**Fig 1: The front end of the Case Tool with the Welcome Message from the Fast Interconnections Group.**

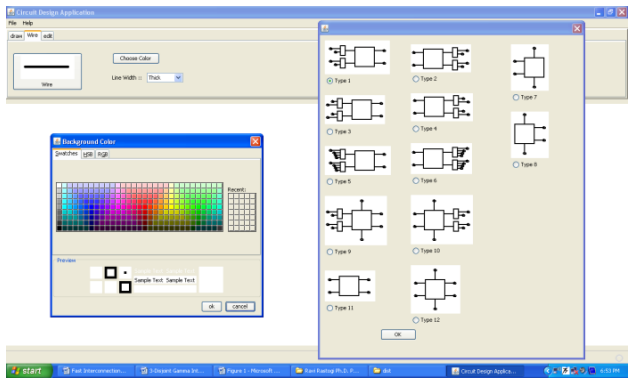

**Fig 2: The front window with different widths of the wires, MIN Components and Color Chooser Applet.**

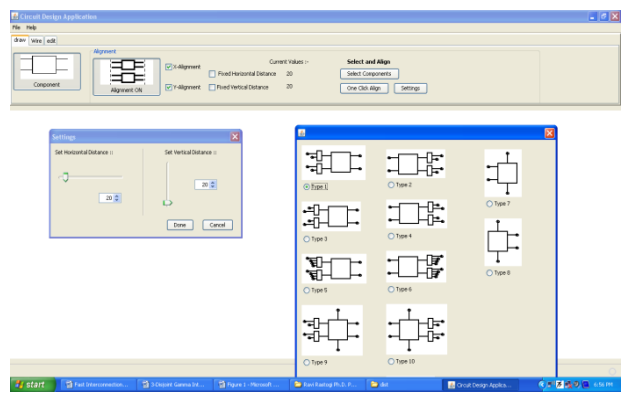

**Fig 3: A case tool with various components and size of the application window can be fixed in terms of horizontal and vertical distance.**

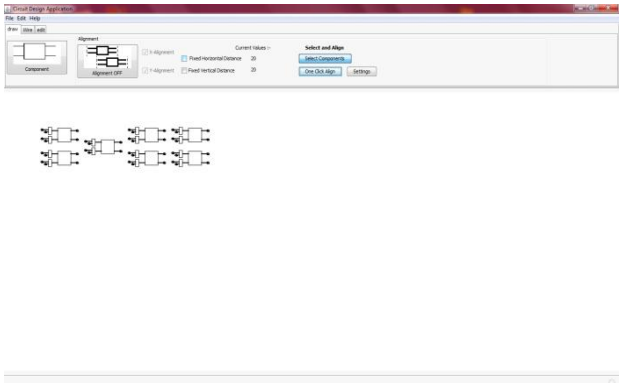

**Fig 4: Shows that the elements have been aligned.**

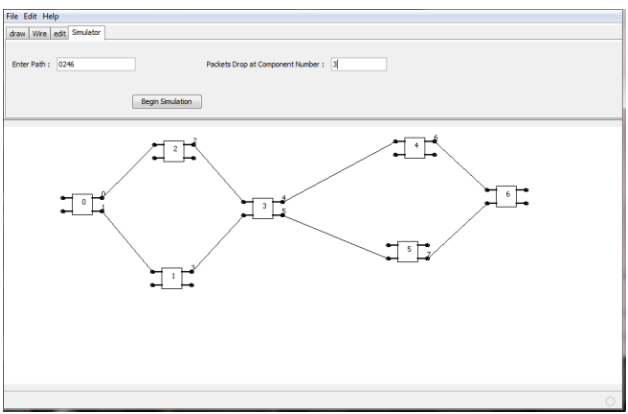

**Fig 5: Shows that the components drawn using the draw method. We have changed the draw method which we have presented last time because of the addition of the new module.**

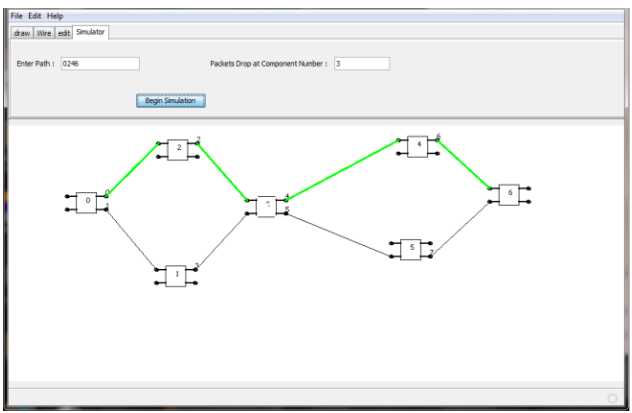

**Fig 6: Highlighting the path in the MIN.**

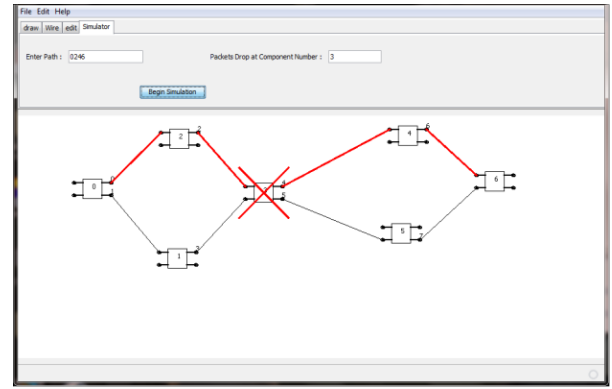

**Fig 7: MIN with one faulty Component and dropping of Packet.**

### 2.5 Code for the New Module

```
@Action
public Task simulate() throws Exception
{
 sim_paint();
 return new SimulateTask(getApplication());
}

public void load_sim(File file) throws Exception
{
 this.path = file.getPath();;
 System.out.println(this.path);
 ois = new ObjectInputStream(new
         FileInputStream(this.path));
for(int i = 0;i < 100;i++)
{
 cmp[i] = (cmp1)ois.readObject();
 lc[i] = (Color)ois.readObject();
 thick[i] = (Integer)ois.readObject();
 line2[i] = (Integer)ois.readObject();
}
for(int i = 0;i < 100;i++)
{
 for(int j = 0;j < 4;j++)
 {
  line[i][j] = (Integer)ois.readObject();
 }
}
 no_cmp = (Integer)ois.readObject();
 no_line = (Integer)ois.readObject();
 ois.close();
 redraw();
}
void sim_paint()
{
 int pkt=0;
 Point p1,p2,p3,p4;
 Graphics2D g=(Graphics2D)
canvas1.getGraphics();
 g.setStroke(new BasicStroke());
 g.setColor(Color.BLACK);
 int x=0;
 String a=jTextField1.getText();
 int[] path1=new int[a.length()];
 for(int i=0;i<a.length();i++)
```

```
 {
path1[i]=Character.getNumericValue(a.charAt(i));
 }
 String b=jTextField2.getText();
 int[] cmp1=new int[b.length()];
 for(int i=0;i<b.length();i++)
 {
cmp1[i]=Character.getNumericValue(b.charAt(i));
 }
 int flag1=0;
 for (int i=0;i<a.length();i++)
 {
  if(path1[i]>no_line)
  {
   flag1=1;
   break;
  }
 }
int flag2=0;
for (int i=0;i<b.length();i++)
{
 if(cmp1[i]>no_cmp)
 {
  flag2=1;
  break;
 }
}
int time=5000000;
if(flag1==1 || flag2==1)
{
 time=0;
 if(flag1==1)
JOptionPane.showMessageDialog(this.canvas1,
"Invalid Path. Please check the input.");
 else if(flag2==1)
JOptionPane.showMessageDialog(this.canvas1,
"Invalid Component number. Please check the
input.");
}
for(int i=0;i<time;i++)
{
 int j;
 if(i%(time/125)<24000)
 {
  g.setColor(Color.RED);
 for(j=0;j<b.length();j++)
 {
  p3=cmp[cmp1[j]].centre;
  pkt++;
  g.drawLine(p3.x-40, p3.y-40, p3.x+40,p3.y+40);
  g.drawLine(p3.x-40, p3.y+40, p3.x+40,p3.y-40);
 }
}
else
{
 g.setColor(g.getBackground());
 for(j=0;j<b.length();j++)
 {
  p3=cmp[cmp1[j]].centre;
   g.drawLine(p3.x-40, p3.y-40, p3.x+40,p3.y+40);
   g.drawLine(p3.x-40, p3.y+40, p3.x+40,p3.y-40);
 }
 g.setColor(Color.GREEN);
}
for(j=0;j<a.length();j++)
{
 Stroke d= new BasicStroke(3);
 g.setStroke(d);
p1=cmp[line[path1[j]][0]].getPoint(line[path1[j]][1]);
p2=cmp[line[path1[j]][2]].getPoint(line[path1[j]][3]);
 g.drawLine(p1.x,p1.y,p2.x,p2.y);
  }
 }
}
//the paint function has been modified a little
too, //new code is :
void redraw()
{
 Point p1,p2,p3,p4;
 Graphics2D g=(Graphics2D)
canvas1.getGraphics();
 canvas1.paint(g);
 for(int j=0;j<no_cmp;j++)
 {
  cmp[j].draw(g);
  g.drawString(Integer.toString(j),cmp[j].
  centre.x,cmp[j].centre.y);
 }
 g.setStroke(new BasicStroke());
 g.setColor(Color.BLACK);
for(int j=0;j<no_line;j++)
{
 Stroke d= new BasicStroke(thick[j]);
 g.setStroke(d);
 g.setColor(lc[j]);
if(line2[j]!=1)
{
 p1=cmp[line[j][0]].getPoint(line[j][1]);
 p2=cmp[line[j][2]].getPoint(line[j][3]);
 g.drawLine(p1.x,p1.y,p2.x,p2.y);
 g.drawString(Integer.toString(j),p1.x ,(p1.y-2)
);
}
else
{
 p1=cmp[line[j][0]].getPoint(line[j][1]);
 p2=cmp[line[j][0]].getPoint(line[j][1]);
 p3=cmp[line[j][2]].getPoint(line[j][3]);
 p4=cmp[line[j][2]].getPoint(line[j][3]);
 int wtm1,wtm2;
 if(cmp[line[j][0]].top_bot(line[j][1])==1)
  wtm1=2*cmp[line[j][0]].getW();
 else
  wtm1=(int) (1.5 * cmp[line[j][0]].getW());
 if(cmp[line[j][2]].top_bot(line[j][1])==1)
  wtm2=2*cmp[line[j][2]].getW();
 else
```

```
wtm2=(int) (1.5 *cmp[line[j][2]].getW());
if(p1.x<=p3.x)
{
 g.drawLine(p1.x,p1.y,p2.x+wtm1,p2.y);
 g.drawLine(p2.x+wtm1,p2.y,p3.x+wtm2,p3.y);
 g.drawLine(p3.x+wtm2,p3.y,p4.x,p4.y);
}
else
{
g.drawLine(p1.x,p1.y,p2.x-wtm1,p2.y);
g.drawLine(p2.x-wtm1,p2.y,p3.x-wtm2,p3.y);
g.drawLine(p3.x-wtm2,p3.y,p4.x,p4.y);
 }
  }
 }
}
```

## 3. CONCLUSION AND FUTURE WORK

In this paper, we have discussed a the newly added module to the existing tool called as Fast Interconnections, which have been designed to develop the 2 and 3-disjoint path multi-stage interconnection network. We have provided the algorithm of the new simulator supported by the screen shots and pseudocode.

The current of the newly added module are as follows- simulation run time and the amount of packets dropped is currently fixed, faulty components are to be input before simulation starts. Further work- we will maintain a database for dropped packets, and allow user to dynamically drop packets from anywhere in the circuit during simulation.